\documentclass[a4paper]{article}
\usepackage{RR}
\RRNo{6088}
\RRdate{Janvier 2007}
\RRversion{2}

\usepackage[english,french]{babel}
\usepackage{multirow}         
\usepackage{array}            
\usepackage{graphicx}          
\usepackage{float}            
\usepackage{amsfonts}         
\usepackage{amssymb}          
\usepackage{amsmath}
\usepackage{makeidx}           
\usepackage[pagebackref]{hyperref} 
\usepackage{algorithmicx}
\usepackage[ruled]{algorithm}
\usepackage{algpseudocode}
\usepackage{amsthm}

\newtheorem{lemma}{Lemma}
\newtheorem{theorem}{Theorem}

\newtheorem{definition}{Definition}
\newtheorem{remark}{Remark}

\newtheorem{property}{Property}

\newcounter{saveline1}

\RRauthor{Pierre Sens \and Luciana Arantes \and Mathieu Bouillaguet
\and V\'eronique Martin \thanks[sfn]{LIP6 - University  of Paris 6 -
INRIA} \and Fab\'iola Greve \thanks{DCC - Computer Science
Department / Federal University of Bahia}}

\authorhead{Sens \& Arantes \& Bouillaguet \& Simon  \& Greve
}

\RRtitle{Impl\'ementation asynchrone de d\'etecteurs de fautes sans
  conna\^itre les participants et  en pr\'esence d'une connectivit\'e partielle}

\RRetitle{Asynchronous Implementation of Failure Detectors with partial
  connectivity and unknown participants}

\titlehead{Failure detector for dynamic networks}

\RRresume{ Cet article aborde le probl\`eme de la d\'etection de
fautes dans les r\'eseaux dynamiques de type MANET. Les d\'etecteurs
de fautes non fiables fournissent des informations sur les processus
d\'efaillants. Ils permettent de r\'esoudre le consensus dans les
r\'eseaux asynchrones. Cependant, la plupart des d\'etecteurs
consid\`ere un ensemble connu de processus interconnect\'es par un
r\'eseau compl\`etement maill\'e. Des telles hypoth\`eses ne sont
pas r\'ealistes dans les environnements dynamiques.
G\'en\'eralement, les impl\'ementations des d\'etecteurs reposent
sur des temporisateurs dont les bornes sont particuli\`erement
difficiles \`a d\'eterminer dans le contexte des r\'eseaux
dynamiques. Cet article pr\'esente une impl\'ementation asynchrone
de d\'etecteurs de d\'efaillances adapt\'ee aux environnements
dynamiques. Nous prouvons que notre algorithme permet
d'impl\'ementer un d\'etection de classe $\diamondsuit S$ lorsque
des propri\'et\'es sur la vitesse relative des transmissions et la
connectivit\'e sont satisfaites par le r\'eseau sous-jacent.}

\RRabstract{ We consider the problem of failure detection in dynamic
networks
  such as MANETs. Unreliable failure detectors are classical
  mechanisms which provide information about process failures.
  However, most of current implementations consider that the network
  is fully connected and that the initial number of nodes of the
  system is known. This assumption is not applicable to dynamic
  environments. Furthermore, such implementations are usually
  timer-based while in dynamic networks there is no upper bound for
  communication delays since nodes can move. This paper presents an
  asynchronous implementation of a failure detector for unknown and mobile networks.
Our approach does not rely on timers and neither the composition nor
the number of nodes in the system are known. We prove that our
algorithm can implement failure detectors of class $\diamondsuit S$
when behavioral properties and connectivity conditions are satisfied
by the underlying system. }

\RRmotcle{détecteurs de fautes, algorithmes répartis, réseaux dynamiques
}
\RRkeyword{failure detectors, distributed algorithms, dynamic networks
}

\RRprojets{Regal}

\RRtheme{\THCom}

\URRocq

\date{}

\begin{document}

\makeRR

\section{Introduction}


The distributed computing scenario is rapidly evolving for
integrating unstructured, self-organizing and dynamic systems, such
as peer-to-peer, wireless sensor and mobile ad-hoc networks.
Nonetheless, the issue of designing reliable services which can
cope with the high dynamism of these systems is a challenge.

Failure detector is a fundamental service, able to help in the
development of fault-tolerant distributed systems. Its importance has
been revealed by Chandra and Toueg who proposed the abstraction of
unreliable failure detectors in order to circumvent the
impossibility result of the consensus problem in an asynchronous
environment~\cite{fischer85impossibility, chandra96unreliable}.
{\it Unreliable failure detectors}, namely FD, can informally be
seen as a per process oracle, which periodically provides a list of
processes suspected of having crashed. In this paper, we are
interested in the class of FD denoted $\diamondsuit S$. Chandra and
Toueg proved that by adding FD of class $\diamondsuit S$ to an
asynchronous system, it is possible to deterministically solve the
consensus problem (with the additional assumption that a majority of
processes are correct).
This paper focuses on FD for mobile and unknown networks, such as
mobile ad-hoc networks (MANETs).
This kind of network presents the following properties: (1) a node
does not necessarily know all the nodes of the network. It can only
send messages to its neighbors, i.e., those nodes that are within
its transmission range~\footnote{The concept of range models, for
instance, homogeneous radio communication in MANETs.};
(2) message transmission delay between nodes is highly
unpredictable; (3) the network is not fully connected which means
that a message sent by a node might be routed through a set of
intermediate nodes until reaching the destination node; (4) a node
can move around and change its transmission range.

Most of current implementations of failure detectors are based on an
all-to-all communication approach where each process periodically
sends a {\it heartbeat} message to all processes
\cite{larrea00,sotoma:adaptation,chandra:implementation}. As they
usually consider a fully connected set of known nodes, these
implementations are not adequate for dynamic environments for the
reasons explained above.  Furthermore, they are usually timer-based,
assuming that eventually some bound of the transmission will
permanently hold.  Such an assumption is not suitable for dynamic
environments where communication delays between two nodes can vary
due to mobility of nodes. In \cite{mostefaoui03async}, Mostefaoui
{\it et al.} have proposed an asynchronous implementation of FDs
which is timer-free. It is based on an exchange of messages which just uses the value of $f$ (the maximum number of processes that can crash) and $n$ (the number of nodes in the system). However, their computation model consists of a set of fully connected initially known nodes.
Some recent works have been proposed which deals with the scalable nature of dynamic systems~\cite{larrea00, gupta01scalable, BMS03}. Nonetheless, few of them tolerate mobility of nodes~\cite{fri05,tai04} and they are all timer-based.

This paper presents a new asynchronous FD algorithm for dynamic systems of mobile and unknown networks. It does not rely on timers to detect failures and no knowledge about the system composition nor its cardinality is required. Yet, it has some interesting features that allow for scalability. The detection of process failures is based only on a local perception that the node has on the network and not on global exchanged information. 

The basic principle of our FD is the flooding of failure suspicion information over the network.
Initially, each node only knows itself. Then, it periodically exchanges a {\sc query-response} pair of messages with its neighbors, that is, those nodes from which it has received a message previously. Then, based only on the reception of these messages and the partial knowledge about the system membership (i.e., its neighborhood), a node is able to suspect other processes or revoke a suspicion in the system. This information about suspicions and mistakes is piggybacked in the {\sc query} messages. Thus, as soon as the underlying system satisfies an {\it f-covering property}, suspicions and mistakes are propagated to the whole network. 
The {\it f-covering} property ensures that there is always a path between any two nodes of the network, in spite of $f$ faults ($f < n$). 

Moreover, if the processes in the system satisfy some behavioral properties, our algorithm implements the failure detectors properties of the class $\diamondsuit S$. Four behavioral properties have been defined.  
The {\it membership property} states that, in order to be known in the system, a node should interact (by sending messages) at least once with some others. The {\it mobility property} states that a moving node should reconnect to the network longtime enough in order to update its state regarding failure suspicions and mistakes. The {\it responsiveness property}  states that after a given time, communication between some node in the system and its neighborhood is always faster than the other communications of this neighborhood. Finally, the {\it mobility responsiveness property} states that at least one correct node in the system does satisfy the responsiveness property and that  its neighborhood is composed of non-moving nodes. 



The rest of the paper is organized as follows. Section \ref{fd}
presents Chandra-Toueg's failure detectors. Section \ref{model}
defines the computation model. In Section \ref{algo}, our
asynchronous failure detector algorithm is presented considering
that nodes do not move.
Section \ref{mobility} describes how the algorithm can be extended to support mobility of nodes. Simulation performance results are shown in Section \ref{perf}, while some related work are
briefly described in Section \ref{relwork}. Finally, Section \ref{conclusion} concludes the paper.

\section{Chandra-Toueg's Failure Detectors}
\label{fd}

Unreliable failure detectors provide information about the aliveness
of processes in the system~\cite{chandra96unreliable}.  Each process has access to a
local failure detector which outputs a list of processes that it
currently suspects of having crashed.  
The failure detector is {\it unreliable} in the sense that it may erroneously add to its
list a process which is actually correct. But if the detector later
believes that suspecting this process is a mistake, it then removes
the process from its list. Therefore, a detector may repeatedly add
and remove the same process from its list of suspected processes.

Failure detectors are formally characterized by two properties.
{\it Completeness} characterizes its capability of suspecting every
faulty process permanently. {\it Accuracy} characterizes its capability of not suspecting correct processes.
Our work is focused on the class of {\it Eventually Strong }
detectors, also known as $\diamondsuit S$. This class contains all
the failure detectors that satisfy (1)
{\small\sf Strong completeness}: there is a time after which every
process that crashes is permanently suspected by every correct
process; (2) {\small\sf Eventual weak accuracy}: there is a time
after which some correct processes are not suspected by any correct
process.

\section{Model}
\label{model}


We consider a dynamic distributed system consisting of a finite set
$\Pi$ of $n > 1$ mobile nodes, namely, $\Pi=\{p_1,\ldots,p_{n}\}$.
Contrarily to a static environment, in a dynamic system of mobile unknown networks, processes
are not aware about $\Pi$ and its cardinality $n$. Thus, they know
only a subset of processes in $\Pi$. There is one process per node
and they communicate by sending and receiving messages via a packet radio network.
There are no assumptions on the relative speed of processes or on
message transfer delays, thus the system is asynchronous. A process
can fail by crashing. A {\it correct} process is a process that does
not crash during a run; otherwise, it is {\it faulty}.
Let $f$ denote the maximum number of processes that may
crash in the system ($f < n$). We assume that $f$ is known to every
process.
To simplify the presentation, we take the range
$\mathcal{T}$ of the clock's tick to be the set of natural numbers.
Processes do not have access to $\mathcal{T}$: it is introduced for
the convenience of the presentation.

The system can be represented by a communication graph $G(V, E)$ in
which $V \subseteq \Pi$ represents the set of nodes and $E$
represents the set of logical links.
Nodes $p_i$ and $p_j$ are connected by a link $(p_i, p_j) \in E$ iff they are within
their wireless transmission range.
In this case, $p_i$ and $p_j$ are considered $1$-hop {\em neighbors}, belonging to the same {\em neighborhood}.
The topology of $G$ is dynamic.
Links are considered to be reliable: they do not create, alter or lose messages. Then, a message $m$ broadcast by $p_i$ is heard by all correct processes in $p_i$'s neighborhood.
Communications  between $1$-hop neighbors are either broadcast or point-to-point.

When a node moves, we consider that it is separated from $G$.
Afterwards, when it stops moving and reconnects to the network, it
is reinserted to $G$. 
A node can keep continuously moving and reconnecting, or eventually it crashes. Nonetheless, a correct moving node will always reconnect to the network. A {\em moving} node is
one that is separated from $G$ and a {\em non-moving} node is
connected to $G$. Let $p_m$ be a moving node. We consider that $p_m$
is not aware about its mobility. Thus, it cannot notify its
neighbors about its moving. In this case, for the viewpoint of a
neighbor, it is not possible to distinguish between a moving or a
crash of $p_m$. During the moving, $p_m$ keeps its state, that is,
the values of its variables.

\begin{definition}{\bf Range:} In a network represented by $G(V, E)$, $range_i$ includes $p_i$ and the set of its 1-hop neighbors. In this case, $|range_i|$ is equal to the degree of $p_i$ in $G$ plus
1. Note that $ranges$ are symmetric i.e. $p_i \in range_j \Rightarrow p_j \in range_i$
\end{definition}


\begin{definition}{\bf Range Density:} In a network represented by $G(V, E)$, the range density, namely $d$, is equal to the size of the smallest range set of the network:

\hspace{1.5 cm}
$d \stackrel{def}{=} min(|range_i|), \forall p_i \in \Pi $

\end{definition}

We assume that $d$ is known to every process.


\begin{definition} {\bf f-Covering Network:}
A network represented by $G(V, E)$ is {\it f-covering} if and only if $G$ is $(f+1)$-connected.
\end{definition}

By Menger's Theorem\cite{yellen98}, a graph $G$ is $(f+1)$-connected if and only if
it contains $(f+1)$ independent paths between any two nodes.
Thus, removing $f$ nodes from $G$ leaves at least {\em one} path between any pair of nodes ($p_i, p_j$).
Moreover, the range density $d$ of the network will be greater than $f + 1$, $d > f + 1$.
These lead to the following remark.
\begin{remark}
Let $G(V, E)$ be an {\it f-covering network}, thus there is a path between any two nodes in $G$, in spite of $f < n$ crashes.
\end{remark}



\section{Implementation of a Failure Detector of Class $\diamondsuit
S$} \label{algo}
This section presents a failure detector algorithm for a network where nodes do not move. The next section (\ref{mobility}) extends this algorithm to support node mobility.
This section firstly presents the principle of the {\em
query-response} mechanism on which our algorithm is based. Then, it
introduces some behavioral properties that, when satisfied by the
underlying system, allow to implement a failure detector of the
class $\diamondsuit S$. Based on such a properties, we propose an
asynchronous failure detection algorithm. A proof that our
implementation provides a failure detector of class $\diamondsuit S$
is also presented.

\subsection{Query-Response Mechanism}
The basic principle of our approach is the flooding of failure
suspicion information over the network based on a local query-response mechanism.
The algorithm proceeds execution by rounds.
At each query-response round, a node broadcasts a {\sc query}
message to the nodes of its range until it possibly crashes. The
time between two consecutive queries is finite but arbitrary. A {\sc
query} message sent by a node includes two sets of nodes: the set
of nodes that it currently suspects of being faulty, and a set of the
mistakes i.e., the nodes that were erroneously suspected of being
faulty previously. Each node keeps a counter, which is incremented
at every round.  Every new information that is generated by this
node about failure suspicions or correction of false suspicions
(mistakes) within a round is tagged with the current value of such a
counter. This tag mechanism avoids old information to be taken into
account by nodes of the network.

Upon receiving a {\sc query} message from a node of its range, a
node sends it back a {\sc response} message. A \textsc{query} issued by a node is satisfied when it receives at least $d-f$ corresponding {\sc response} messages.
Moreover, each couple of {\sc query}- {\sc response} messages are uniquely
identified in the system\footnote{For the sake of simplicity, such
identification is not included in the code of the algorithms of the
paper.}.
Notice that we assume that a node issues a new \textsc{query} only after the
previous one is terminated. Moreover, when a node broadcasts a
\textsc{query} message, we assume that it receives the \textsc{query} too, and that
its own response always arrives  among the first $d-f$ responses it
is waiting for.

\subsection{Behavioral Properties}

Let us define some behavioral properties that processes should have
in order to ensure that our proposed implementation of a failure
detector satisfies the properties of class $\diamondsuit S$ in an
unknown network.

In order to implement any type of unreliable failure detector with
an unknown membership, processes should interact with some others to
be known. According to~\cite{FernandezPRDC06}, if there is some
process in the system such that the rest of processes have no
knowledge whatsoever of its identity, there is no algorithm that
implements a failure detector with weak completeness, even if links
are reliable and the system is synchronous. Thus, in order to
implement a $\diamondsuit S$ failure detector, the following {\it
membership property}, namely $\mathcal{MP}$, should be ensured by
all processes in the system.




\begin{property} {\bf Membership Property ($\mathcal{MP}$)}.  Let $t
  \in \mathcal{T}$. Denote $known_j^t$ the set of processes from which
  $p_j$ has received a {\sc query} message at time $t$.  Let $K_i^t$
  be the set of processes $p_j$ that, at time $t$, have received a
  {\sc query} from $p_i$. That is, $K_i^t = \{p_j ~ | ~ p_i \in
  known_j^t \}$.  A process $p_i$ satisfies the membership property
  if:

\hspace{1.0cm}
$ \mathcal{MP}(p_i) \stackrel{def}{=} \exists t \geq 0 \in \mathcal{T}: | K_i^t | > f + 1$
\end{property}

This property states that, to be part of the membership of the system, a process $p_i$ (either correct or not) should interact at least once with other processes in its range by broadcasting a {\sc query} message. Moreover, this query should be received and represented in the state of at least one correct process in the system, beyond the process $p_i$ itself.

Let us define another important property in order to implement a timer-free failure detector in a system with an unknown membership. It is the {\it responsiveness property}, namely $\mathcal{RP}$, which denotes the ability of a node to reply to a {\sc query} among the first nodes.

\begin{property} {\bf Responsiveness Property ($\mathcal{RP}$)}.  Let
  $t, u \in \mathcal{T}$. Denote $rec\_from_j^t$ the set of $d - f$
  processes from which $p_j$ has received responses to its {\sc query}
  message that terminated at or before $t$.
  The $\mathcal{RP}$ property of the correct process $p_i$ is defined as follows:

  $ \mathcal{RP}(p_{i}) \stackrel{def}{=} \exists u \in \mathcal{T}:
  \forall t > u, \forall p_j \in range_i,\, p_i \in rec\_from_j^t $

\end{property}

Intuitively, the $\mathcal{RP}(p_{i})$ property states that after a
finite time $u$, the set of the $d-f$ responses received by any
neighbor of $p_i$ to its last \textsc{query} always includes a response from $p_i$.


\subsection {Implementation of a Failure Detector of Class $\diamondsuit S$ for Unknown Networks}

Algorithm \ref{alg:FD_async} describes our protocol for implementing
a failure detector of class $\diamondsuit S$ when the underlying
system is an {\it f-covering} network, satisfying the behavioral properties.



We use the following notations:

\begin{itemize}
  \item $counter_i$: denotes the round counter of node $p_i$.

\item $suspected_i$: denotes the current set of processes suspected of
  being faulty by $p_i$. Each element of this set is a tuple of the form $\langle id, counter \rangle$,
  where $id$ is the identifier of the suspected node and $counter$ is
  the value of $counter_i$ when $p_i$ generated the information that it suspected node $id$ of being faulty.

\item $mistake_i$: denotes the set of nodes which were
previously suspected of being faulty but such suspicions are
currently considered to be false. Similar to the $suspected_i$ set,
the $mistake_i$ is composed of tuples of the form $\langle
id,counter \rangle$ i.e, $counter$ indicates when the information
that $id$ is falsely suspected was generated.


\item $rec\_from_i$: denotes the set of nodes from which $p_i$ has
  received responses to its last {\sc query} message.


\item $known_i$: denotes the current knowledge of $p_i$ about its neighborhood. $known_i$ is then the set of processes from
  which $p_i$ has received a {\sc query} messages since the beginning of execution.


\item $Add(set,\langle id,counter \rangle)$: is a function that includes
  $\langle id,counter \rangle$ in $set$. If an $\langle id, - \rangle$ already exists in $set$, it
  is replaced by $\langle id,counter \rangle$.

\end{itemize}



\begin{algorithm}[h]
 \caption{Asynchronous Implementation of a Failure Detector}
 \label{alg:FD_async}

   \begin{algorithmic}[1]
     \footnotesize
     \State \textbf{init:}
     \State $suspected_i \gets \varnothing ; mistake_i \gets \varnothing; counter_i \gets
     0$
     \State $known_i \gets \varnothing$


     \Statex \State \textbf{Task} T1:
     \Loop

     \State broadcast {\sc query}($suspected_i$, $mistake_i$) \label{broadcast}
     \State \textbf{wait until} {\sc  response} received from at least $(d-f)$ distinct
     processes \label{receive}

     \State $rec\_from_i \gets$  the set of distinct nodes from which $p_i$ has received a response at line
     \ref{receive} \label{recfrom}

      \For{\textbf{all} $p_j \in known_i \setminus rec\_from_i \mid
        \langle  p_j, -  \rangle \not\in suspected_i$} \label{new_susp}
        \If {$\langle p_j, counter \rangle \in mistake_i$}
           \State $counter_i = max(counter_i, counter +1)$
           \label{update_counter1_T1}
            \State $mistake_i = mistake_i \setminus \langle p_j, -
            \rangle$ \label{del_mist}
        \EndIf
        \State $Add(suspected_i, \langle p_j, counter_i \rangle)$
        \label{add_susp}

       \EndFor \label{end_new_susp}

       \State $counter_i = counter_i + 1$ \label{update_counter2_T1}
     \EndLoop

     \Statex \State \textbf{Task} T2:
     \State \textbf{upon reception of} {\sc query}
     ($suspected_j$,$mistake_j$) \textbf{from} $p_j$ \textbf{do} \label{receive_query}

     \State $known_i \gets known_i \cup \{p_j\}$ \label{update_known_set}

     \For{ \textbf{all} $\langle p_x,counter_x \rangle \in
       suspected_j$} \label{forw_susp}

       \If {$\langle p_x, - \rangle \not \in suspected_i \cup mistake_i$  \textit{or}
       $\langle p_x, counter \rangle \in suspected_i \cup mistake_i  \mid counter <
       counter_x$}\label{suspect_condition}

              \If {$p_x = p_i$} \label{new_mistake}
                \State $counter_i = max(counter_i, counter_x +1)$ \label{update_counter_mistake}
                \State $Add(mistake_i, \langle p_i, counter_i \rangle)$ \label{add_new_mistake}
              \Else \label{end_new_mistake}

                 \State $Add(suspected_i,\langle p_x,counter_x \rangle)$ \label{add_suspected}
                  \State $mistake_i = mistake_i \setminus \langle p_x, -
                  \rangle$ \label{remove_mistake}
               \EndIf
       \EndIf
       \EndFor \label{end_forw_susp}

       \For{ \textbf{all} $\langle p_x,counter_x \rangle \in
         mistake_j$} \label{forw_mistake}

       \If {$\langle p_x, - \rangle \not \in suspected_i \cup
         mistake_i$ \textit{or}  $\langle p_x, counter \rangle \in
        suspected_i \cup mistake_i \mid counter \leq counter_x$} \label{mistake_condition}
       \State $Add(mistake_i,\langle p_x,counter_x \rangle)$ \label{add_mistake}
       \State $suspected_i = suspected_i \setminus \langle p_x, -  \rangle$ \label{remove_susp}
       \setcounter{saveline1}{\value{ALG@line}}

       \EndIf

       \EndFor \label{end_forw_mistake}

     \State send {\sc response} \textbf{to} $p_j$
     \label{send_response}
   \end{algorithmic}
 \end{algorithm}

The algorithm is composed of two tasks.  Task $T1$ is made up of an
infinite loop. At each round, a {\sc query} message is sent to all
nodes of $p_i$'s range (line \ref{broadcast}). This message includes
the set of nodes that $p_i$ currently suspects and the set of
mistakes of which $p_i$ is aware.  Node $p_i$ waits for at least
$d-f$ responses, which includes $p_i$'s own response (line
\ref{receive}).
Then, $p_i$ detects new suspicions (lines
\ref{new_susp}-\ref{end_new_susp}). $p_i$ starts suspecting each non
previously suspected node $p_j$ that it knows ($p_j \in known_i$)
but from which it does receive a {\sc response} to its last {\sc
query}. If a previous mistake information related to this new
suspected node exists in the mistake set $mistake_i$, it is removed
from it (line \ref{del_mist}) and the counter $counter_i$ is updated
to a value greater than the mistake tag (line
\ref{update_counter1_T1}). The new suspicion information is then
included in $suspected_i$ with a tag which is equal to the current
value of $counter_i$ (line \ref{add_susp}). Finally, at the end of
task T1, $counter_i$ is incremented by one (line
\ref{update_counter2_T1}).

Task $T2$ allows a node to handle the reception of a {\sc query}
message sent by another node of its range. A {\sc query} message
contains the information about suspected nodes and mistakes kept by
the sending node. However, based on the tag associated to each piece
of information, the receiving node only takes into account the ones
that are more recent than those it already knows.

The two loops of task $T2$ respectively handle the information
received about suspected nodes (lines
\ref{forw_susp}--\ref{end_forw_susp}) and about mistaken nodes
(lines \ref{forw_mistake}--\ref{end_forw_mistake}). Thus, for each
node $p_x$ included in the suspected (respectively, mistake) set of
the {\sc query} message, $p_i$ includes the node $p_x$ in its
$suspected_i$ (respectively, $mistake_i$) set only if the following
condition is satisfied: $p_i$ received a more recent information
about $p_x$ status (failed or mistaken) than the ones it has in its
$suspected_i$ and $mistake_i$ sets. A more recent information is
characterized by the fact that $p_x$ has never been suspected or
false suspected by $p_i$ or by the fact that its $counter$ in the
$p_i$ sets is less than the new received $counter_x$ (see lines
\ref{suspect_condition} and \ref{mistake_condition}). In such a
case, $p_i$ also removes the node $p_x$ from its $mistake_i$
(respectively, $suspected_i$) set (lines \ref{remove_mistake} and
\ref{remove_susp}).

Furthermore, in the first loop, a new mistake is detected if the
receiving node $p_i$ is included in the suspected set of the {\sc
query} message (line \ref{new_mistake}). Then, $p_i$ adds itself in
its local mistake set (line \ref{add_new_mistake}).  The tag
$counter_i$ associated to this mistake is equal to the maximum  of
the current value of $counter_i$ and the tag associated to the
suspicion of $p_i$, included in $suspected_j$ set, incremented by
one (line \ref{update_counter_mistake}).
At the end of task $T2$ (line \ref{send_response}), $p_i$ sends to
the querying node a {\sc response} message.

\subsection{Example of the Execution of the Algorithm}

\begin{figure}[htbp]
  \centering
  \includegraphics[scale=0.4]{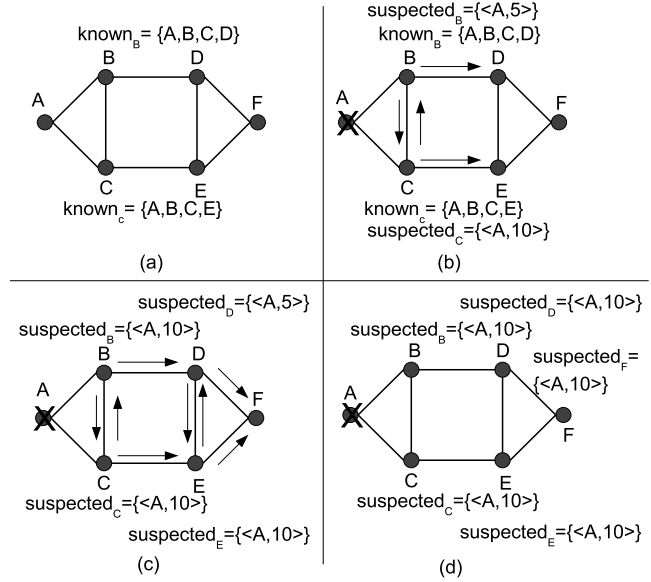}
  \vspace{-0,4cm} \caption{Example of Failure Detection}
  \label{fig:example}
\end{figure}

  Figure \ref{fig:example} illustrates an execution which shows the strong completeness property of Algorithm
  \ref{alg:FD_async}. We consider an {\it 1-covering} network ($f=1$) whose range
  density is equal to 3. Thus, each querying node should wait for at least 2 responses (one from itself and the other from one of its neighbors).

We do not show a scenario from the beginning of execution of the
algorithm, but one where every node $i$ is already aware of the
participants of its range ($known_i$), see step (a). In step (b),
$A$ fails.  Thus, as neither node $B$ nor node $C$ receive a
responses from $A$ to their respective \textsc{query}, they start
suspecting $A$.  At the moment of the \textsc{query} $counter_B$ is
equal to $5$ (see $suspected_{B}$) but $counter_C$ is equal to $10$
(see $suspected_{C}$). Then, both $B$ and $C$ propagate their
suspected sets to their neighbors in their next respective {\sc
query} rounds as shown in step (c). Nodes $D$ and $E$ will include
the corresponding information $\langle A,5 \rangle$ and $\langle
A,10 \rangle$  in their respective sets $suspected_D$ and
$suspected_E$. Node $B$ will update its $suspected_B$ set since the
counter of the received information from $C$ is greater than the one
that it keeps in its $suspected_B$. However, $C$ will discard the
information received from $B$. Similar to step (c), in step (d)
nodes $B$, $C$, $D$ and $E$ include in their next {\sc query}
message their respective suspected sets.  Therefore, eventually the
information $\langle A,10 \rangle$ related to the failure of $A$
will be delivered to all correct nodes of the network.

\subsection {Proof}

We present in this section a sketch of proof of both the strong
completeness and eventual weak accuracy properties of our algorithm
that characterize failure detectors of class $\diamondsuit S$
for an {\it f-covering network} composed of non-moving nodes.



Consider that the {\em most recent status about a process} $p_x$ is stored in a $supected$ or $mistake$ set and represented by the tuple $\langle p_x, ct_x \rangle$ which has the greatest counter $ct_x$ in the network. In case of equality between a suspicion and a mistake, we give arbitrarily precedence to the mistake.

\begin{lemma}
  \label{propagation_lemma}
  Consider an {\it f-covering network}.
  Let $p_i$ be a correct process. Consider that, at time $t$, $p_i$ owns the most recent status
  about $p_x$ in the network ($\langle p_x,ct_x \rangle$) in its $suspected_i$ set (respectively, $mistake_i$ set).
   If no more recent information about $p_x$ status is generated afterward, then eventually all correct nodes will
  include $\langle p_x, ct_x \rangle$ in their suspected set (respectively, mistake set).
 \end{lemma}
\begin{proof}
  Since $p_i$ is correct, it will execute line \ref{broadcast} and broadcast a \textsc{query}
  message containing $\langle p_x, ct_x \rangle$ in the $suspected_i$ set (respectively, $mistake_i$ set) to all its neighbors.  As channels are reliable, this $\textsc{query}$ message is received by every correct process $p_j \in range_i$. Thus, $p_j$ will execute lines~\ref{forw_susp}-\ref{end_forw_susp} (respectively, lines \ref{forw_mistake}-\ref{end_forw_mistake}). Since $ct_x$ is the greatest counter associated with $p_x$ in the network, $p_j$ executes line \ref{add_suspected} (respectively, line~\ref{add_mistake}) and add $\langle p_x, ct_x \rangle$ to its own $suspected_j$ set (respectively, $mistake_j$ set).
In the next round, $p_j$, the same as $p_i$, must broadcast this new status regarding $p_x$ in its respective sets.
Thus, due to the  \textit{f-covering} network property, all nodes in the network eventually add $\langle p_x, ct_x \rangle$ in their suspected set (respectively, mistake set) and the lemma follows.
\end{proof}




\begin{lemma}
  \label{L-strong-complet}
  Consider an {\it f-covering network} in which all processes satisfy $\mathcal{MP}$.
  Let $p_f$ be a faulty process. If process $p_i$ is correct then
  eventually $p_f$ is permanently included in its $suspected_i$ set.
\end{lemma}
\begin{proof}
 Let us consider that $p_f$ crashes at time $t$.

 {\small\sf Remark 1}.  Since $\mathcal{MP}(p_f)$ is satisfied, $p_f$ has sent to processes in
  $range_f$ at least one \textsc{query} message before it crashed at time $t$.
  Then, a number of correct processes within $range_{f}$ will include $p_f$ in
  their respective $known$ set which is updated when a
  process receives a \textsc{query} (line \ref{update_known_set}).
  Let us denote $K$ this set of processes.
  Notice that, by $\mathcal{MP}$, $|K| > f + 1$, and then there is at least one correct process $p_i$ such that $p_f \in known_i$.

 {\small\sf Remark 2}.  As $p_f$ has crashed, there will be a time $t' > t$ after which all processes in $K$ will
  never receive a {\sc response} message from $p_f$ (i.e., $p_f \notin
  rec\_from$ sets of processes within $K$) (line \ref{receive}).
  Thus, if $p_f$ was not already suspected by these processes (line
  \ref{new_susp}), it will be included in their corresponding
  suspected sets with a tag equal to the current value of their
  respective $counter$ or with a greater tag then the one associated
  with $p_f$ in the mistake set if it was previously in there (line
  \ref{add_susp}).  At this point no more information about $p_f$ can be generated since
  only $p_f$ can generate a mistake about itself (line~\ref{new_mistake}) and only processes in
  $K$ can generate a new suspicion and $p_f$ is already in their
  suspected set. Thus, the most recent information about $p_f$ sent in
  a \textsc{query} message is either (1) a suspicion or (2) a mistake.
  In the first case, following Lemma~\ref{propagation_lemma}, all
  correct processes will eventually include $p_f$ in their respective
  suspected set. Since no new information about $p_f$ is generated,
  $p_f$ is permanently suspected by all correct nodes. In the second
  case, by Lemma \ref{propagation_lemma}, the mistake eventually reach
  a correct process $p_i$ in $K$, which removes $p_f$ from
  $suspected_i$. At the next round, $p_i$ will include $p_f$ in
  $suspected_i$ with a greater tag since $p_f \notin rec\_from_i$ and
  $p_f \notin suspected_i$. This information will in turn be
  propagated to all correct processes, following the propagation
  Lemma~\ref{propagation_lemma}. Thus, all correct processes will permanently suspect $p_f$
  since no new information about $p_f$ is generated.
  \end{proof}


\begin{lemma}
  \label{L-weak-accuracy}
  Consider an {\it f-covering network} in which all processes satisfy $\mathcal{MP}$.
  Let $p_i$ be a correct process which satisfies the responsiveness property $\mathcal{RP}(p_i)$.
  There is a time $u$ after which $p_i$ is not included in the $suspected_j$ set of any correct process $p_j$.
\end{lemma}
\begin{proof}
   {\small\sf Remark 1}.  According to $\mathcal{RP}(p_i)$, there is a time $t$ after which
  every process $p_j$ in the neighborhood of $p_i$ receives a
  \textsc{response} message from $p_i$ in reply to their query. Thus,
  after time $t$, $p_i$ is always included in the $rec\_from$ sets of
  all nodes within its $range_i$. Since a process starts being
  suspected only if its reply is not received by one of its neighbor
  (lines \ref{new_susp}-\ref{end_new_susp}), no process adds $p_i$ to
  its suspected set due to a \textsc{query} message sent after time
  $t$.

  {\small\sf Remark 2}. If $p_i$ is not included in any suspected set in the network, clearly
  $p_i$ cannot be suspected anymore. If $p_i$ is included in at least
  one suspected set, there are two cases to consider: the most recent
  piece of information about $p_i$ is either (1) a mistake or (2) a
  suspicion. In the first case, based on Lemma
  \ref{propagation_lemma}, all processes which were suspecting $p_i$
  will eventually execute lines \ref{add_mistake}-\ref{remove_susp}
  upon receiving the propagated mistake and remove $p_i$ from their
  suspected set definitely. In the second case, following Lemma
  \ref{propagation_lemma}, $p_i$ will eventually deliver a
  \textsc{query} message with $p_i$ in the suspected set. This will
  cause $p_i$ to generate a new mistake with a greater tag. This
  mistake will in turn be propagated to all processes, which will
  remove $p_i$ from their suspected set if they were suspecting it.
\end{proof}

\begin{theorem}
  \label{L-strong-class} Algorithm~\ref{alg:FD_async} implements a
  failure detector of class $\diamondsuit S$, assuming an {\it
    f-covering network} of non-moving nodes which satisfies the behavioral properties
  $\mathcal{RP}$, $\mathcal{MP}$ and with $f < n$.
\end{theorem}
\begin{proof}
Consider a correct process $p_i$ and a fault process $p_f$.
 To satisfy the {\small\sf strong completeness} property, we must prove that eventually $p_f$ is permanently included in $suspected_i$ set of $p_i$. This claim follows directly from Lemma~\ref{L-strong-complet}.
  To satisfy the {\small\sf eventual weak accuracy} property, we must prove that there is a time
  $u$ after which $p_i$ is not included in the $suspected_j$ set of any correct process $p_j$.
  This claim follows directly from Lemma~\ref{L-weak-accuracy} and the theorem follows.
\end{proof}



\section{Extension for Mobility Management}
\label{mobility}


In this section we present an extension for Algorithm
\ref{alg:FD_async} that supports mobility of nodes. For such an
extension some new behavioral properties in respect to mobility of
nodes and the underlying system must be defined.

\subsection{Mobility Behavioral Properties}

Let $p_m$ be a moving node. Notice that a node can keep continuously
moving and reconnecting, or eventually crashes. Nonetheless, we
consider that $p_m$ should stay connected to the network for a
sufficient period of time in order to be able to update its state
with recent information regarding failure suspicions and mistakes.
Otherwise,
it would not update its state properly and thus completeness and
accuracy properties of the failure detector would not be ensured.
Hence, in order to capture this notion of  ``sufficient time of
reconnection'', the following {\it mobility property}, namely
$\mathcal{M}obi\mathcal{P}$, has been defined.



\begin{property} {\bf Mobility Property ($\mathcal{M}obi\mathcal{P}$)}.
Let $t \in \mathcal{T}$. Let $Q_i^t$ be the set of processes from
which $p_i$ has received a {\sc query} message that terminated before or at $t$.
A process $p_i$ satisfies the mobility property if:

\hspace{0.5cm} $\mathcal{M}obi\mathcal{P}(p_i) \stackrel{def}{=}
\exists t \geq 0 \in \mathcal{T}: | Q_i^t | > f + 1$
\end{property}

This property should be satisfied by all moving nodes when they
reconnect to the network. Thus, $\mathcal{M}obi\mathcal{P}(p_m)$
ensures that, after reconnecting, there will be a time at which
process $p_m$ should have received {\sc query} messages from at
least one correct process, beyond itself. Since {\sc query} messages
carry the state of suspicions and mistakes in the membership, this
ensures that process $p_m$ will update its state with recent
informations.

We assume also that the {\it membership property} holds for all
moving nodes when they reconnect to the network. Thus,
$\mathcal{MP}(p_m)$ ensures that, after reconnecting, there will be
a time at which process $p_m$ interacts at least once with other
processes in its $range_m$, broadcasting a {\sc query} message which
will be delivered by at least one correct processes in $range_m$,
beyond $p_m$.

Regarding the underlying system behavior, we consider that despite
mobility, the {\it f-covering} property of the network is ensured
and that the range density $d$ of the network does not change.
Moreover, we have extended the $\mathcal{RP}$ property such that
neighbors of a node $p$, which has the $\mathcal{RP}$ property,
eventually stop moving outside $p$'s range. Otherwise, even if $p$
has the $\mathcal{RP}$ property, a moving node would add $p$ in its
$known$ set whenever it belonged to $p$'s range and then it would
suspect $p$ when it moved outside $p$'s range. The extension of
$\mathcal{RP}$ property, namely $\mathcal{M}obi\mathcal{RP}$, is
defined as follows:

\begin{property} {\bf Mobility Responsiveness Property ($\mathcal{M}obi\mathcal{RP}$)}.
Let $t \in \mathcal{T}$. Denote $range_i^t$ the set of processes in
$range_i$ at $t$. A process $p_i$ satisfies the mobility
responsiveness property if:

$\mathcal{M}obi\mathcal{RP}(p_i) \stackrel{def}{=}
\mathcal{RP}(p_i):  \exists u \in \mathcal{T}:
  \forall t~>~u, \forall t'~>~t,p_j \in range_i^t \Rightarrow p_j \in range_i^{t'} $
\end{property}

$\mathcal{M}obi\mathcal{RP}$ should hold for at least one correct
non-moving node.

\subsection{Implementation of a Failure Detector of Class $\diamondsuit S$ for Mobile Unknown Networks}

The extension of the algorithm to support mobility of nodes is based
on the same {\em query-response} principle presented in
Section~\ref{algo}. When a node $p_m$ moves to another range, it
starts being suspected of having crashed by those nodes of its old
range, since it cannot reply to {\sc query} messages from the latter
anymore. Hence, {\sc query} messages that include $p_m$ as a
suspected node will be propagated to nodes of the network.
Eventually, when $p_m$ reconnects to the network, it will receive
such suspicion messages. Upon receiving them, $p_m$ will correct
such a mistake by including itself ($p_m$) in the mistake set of its
corresponding {\sc query} messages. Such information will be
propagated over the network.
On the other hand, $p_m$ will start suspecting the nodes of its old
range since they are in its $known$ set. It then will broadcast this
suspected information in its next {\sc query} message. Eventually,
this information will be corrected by the nodes of its old range,
and the corresponding generated mistakes will spread over the
network, following the same principle.
Notice that, in order to avoid a ``ping-pong" effect between
information about failure suspicions and corrections (mistakes), a
mechanism should be added to the algorithm in order to remove from
$known$ sets those nodes that belong to remote ranges.



In Algorithm \ref{alg:FD_async_mob}, we just show the lines which
need to be included in task $T2$ of Algorithm \ref{alg:FD_async} in
order to support mobility of nodes.
Lines \ref{test_origin}--\ref{end_move} should be added in the
\textbf{if} block of the second loop of task $T2$, just after line
\ref{remove_susp} of Algorithm \ref{alg:FD_async}. They allow
the updating of the $known$ sets of both the moving node $p_m$ and
of those nodes that belong to the original range of  $p_m$. For each
mistake $\langle p_x, counter_x \rangle$ received from a node $p_j$
such that node $p_i$ keeps an old information about $p_x$, $p_i$
verifies whether $p_x$ is the sending node $p_j$. In they are
different, $p_x$ should belong to a remote range $range_x$, such that $p_x
\notin range_i$. Thus, process $p_x$ is removed from the local set
$known_i$.

\begin{algorithm}
 \caption{Asynchronous Implementation of a Failure Detector with Mobility of Nodes}
 \label{alg:FD_async_mob}

   \begin{algorithmic}[1]
     \footnotesize




    \item[]
     \setcounter{ALG@line}{\value{saveline1}}

       \If{($p_x \not = p_j$)} \label{test_origin}
       \State $known_i = known_i \setminus \{p_x\}$ \label{remove_from_known_set}
       \EndIf
       \label{end_move}

   \end{algorithmic}
 \end{algorithm}


\subsection {Proof}
\label{S-proof-mobil}

We present in this section a sketch of proof of both the strong
completeness and eventual weak accuracy properties of the extended
algorithm \ref{alg:FD_async_mob} that characterize failure detectors
of class $\diamondsuit S$ for an {\it f-covering network} composed
of moving and non-moving nodes.

\begin{lemma}
\label{L-known}
(1) Infinitely often, during a run, the $known_i$ set contains either correct processes which are in $range_i$ or faulty processes. Moreover, (2) for every process $p_i$ which satisfies $\mathcal{MP}(p_i)$, then there is a correct process $p_j$, such that $p_i \in known_j$.
\end{lemma}

\begin{proof}

Let us observe that the {\sc query-response} messages are exchanged between processes in the same range. Thus, on the execution of line \ref{update_known_set}, the set $known_i$ is updated when $p_i$ receives {\sc query} messages from other processes in its $range_i$. Beyond line~\ref{update_known_set}, $known_i$ may be updated at lines \ref{test_origin}--\ref{end_move}, in order to remove nodes suspected to be in another range, different from $p_i$'s range. This may happen due to a mobility. Thus, if a process which raised a mistake ($p_x$) is different from the process who carries it ($p_j$), probably $p_x$ does not belong to $range_i$, because otherwise, $p_i$ would have received the mistake by $p_x$ itself. It may happen that $p_x$ was in $range_i$ at some point in time, but due to a move, it has changed to another neighborhood, such that $p_x \not \in range_i$. Wherever the case, process $p_i$ is going to remove $p_x$ from its $known_i$ set and the part (1) of this lemma follows.

Let us prove part (2) of the lemma. Since $\mathcal{MP}(p_i)$ is satisfied, there is at least one correct process $p_k$ which has received a \textsc{query} message from $p_i$ after $p_i$ has connected or reconnected to the network at time $t$. Thus, $p_i \in known_k$. Nonetheless, later, $p_i$ can be removed from $known_k$ by the execution of lines~\ref{test_origin}--\ref{end_move} due to a suspicion of mobility.
But, notice that, since channels are reliable, the {\sc query} from $p_i$ in which $p_i \in mistake_i$ is going to eventually arrive to $p_k$. In this case, two situations can occur. Situation (1). If this {\sc query} is the first one to arrive at $p_k$, it will satisfy the predicate of line~\ref{mistake_condition}, thus lines~\ref{add_mistake}--\ref{remove_susp} are executed, but not lines~\ref{test_origin}--\ref{end_move}. Afterward, when a {\sc query} from a process $p_j$ arrives containing the mistake over $p_i$, and such that $p_i \not = p_j$, then since this mistake has already been taken into account, the predicate of line \ref{mistake_condition} will not be satisfied and lines~\ref{test_origin}--\ref{end_move} are not executed. Thus $p_k$ will not remove $p_i$ from $known_k$ set.
Situation (2). A  {\sc query} from a process $p_j$ is the first one to arrive at $p_k$ containing the mistake over $p_i$, and such that $p_i \not = p_j$. In this case, the predicate of line~\ref{mistake_condition} is satisfied and lines~\ref{test_origin}--\ref{end_move} are executed. Thus $p_k$ removes $p_i$ from $known_k$. Nonetheless, later, a {\sc query} from $p_i$ arrives in which $p_i \in mistake_i$. In this case, process $p_k$ will execute line~\ref{update_known_set} including $p_i$ in $known_k$. Moreover, since this mistake has already been taken into account, the predicate of line \ref{mistake_condition} will not be satisfied and lines~\ref{test_origin}--\ref{end_move} are not executed. Thus $p_k$ will not remove $p_i$ from $known_k$ set. This concludes the proof of part (2).


\end{proof}


\begin{lemma}
  \label{propagation_lemma_mobility}
  Consider an {\it f-covering network} in which all nodes satisfy
  $\mathcal{MP}$ and all moving nodes satisfy
  $\mathcal{M}obi\mathcal{P}$.  Lemma~\ref{propagation_lemma} holds
  for every correct process $p_i$ (moving or non-moving) .
\end{lemma}

\begin{proof}
  The lemma follows directly from Lemma~\ref{propagation_lemma} if
  $p_i$ is a non-moving node.  To take into account moving nodes, we
  should consider two cases.  Case (1). Assume that $p_i$ is a correct
  moving node which has the most recent status about process $p_x$. As
  soon as $p_i$ reconnects to the network at time $t'$, it will
  execute line \ref{broadcast} and broadcast a \textsc{query} message
  to all its neighbors. Since $\mathcal{MP}(p_i)$ holds, $p_i$ is
  correct and channels are reliable, every correct node $p_j \in
  range_i$ receives this \textsc{query} message. Since, $|range_i| > f
  + 1$, there will be at least one correct non-moving node $p_k$ which
  receives this \textsc{query}. Thus, by the same arguments of Lemma
  \ref{propagation_lemma}, the lemma follows.

  Case (2). Assume that $p_i$ is a correct moving node which has not
  yet the most recent status about process $p_x$ and let us consider
  that due to Lemma~\ref{propagation_lemma}, every non-moving node
  has added $p_x$ in its $suspected$ (respectively, $mistake$) set
  before or at time $t$. As soon as $p_i$ reconnects to the network
  at time $t' \geq t$, since $\mathcal{M}obi\mathcal{P}(p_i)$
  is satisfied, $p_i$ will receive {\sc query} messages from at least
  a correct process $p_j$ with the last status of suspicion and
  mistaken informations about $p_x$. Thus, $p_i$ will eventually add
  $p_x$ in its $suspected_i$ (respectively, $mistake_i$) set and the
  lemma follows.

\end{proof}

\begin{lemma}
\label{L-mobil-strong-complet}
Consider an {\it f-covering network} in which all nodes satisfy $\mathcal{MP}$ and all moving nodes satisfy  $\mathcal{M}obi\mathcal{P}$. Let $p_f$ be a faulty process (moving or non-moving). If process $p_i$ (moving or non-moving) is correct then eventually $p_f$ is permanently included in its $suspected_i$ set.
\end{lemma}

\begin{proof}
  If $p_i$ and $p_f$ are non-moving nodes, the lemma follows directly
  from Lemma~\ref{L-strong-complet}. To take into account moving
  nodes, let us assume that $p_i$ is a correct {\em moving} node which
  has the most recent status about process $p_f$.
  Due to Lemma~\ref{L-known} and the same arguments of Lemma~\ref{L-strong-complet}
  (Remark 1), $p_f$ is in the $known$ set of at least one correct
  process in the network. We should consider the following cases.

  Case (1).  Consider that $p_f$ crashes at time $r < t$. Let us
  suppose that $p_i$ is the only correct process such that $p_f \in
  known_i$. Moreover, before broadcasting this information to its
  neighborhood, $p_i$ moves at time $t$. Since $p_i$ keeps its state
  during the moving, $p_f \in suspected_i$. When $p_i$ reconnects to
  the network at time $t'$, due to
  Lemma~\ref{propagation_lemma_mobility}, this information about the
  suspicion of $p_f$ will be propagated to all correct nodes in the
  network. Finally, due to the same arguments of
  Lemma~\ref{L-strong-complet} (Remark 2) and
  Lemma~\ref{propagation_lemma_mobility}, $p_f$ is permanently
  included in every $suspected$ set of a correct process, either
  moving or non-moving.

  Case (2). Consider that $p_f$ crashes at time $s$, $t < s < t'$.
  Suppose that $p_i$ has $p_f$ in its $mistake_i$ when it starts
  moving at time $t$. Since $p_i$ keeps its state during the moving,
  $p_f \in mistake_i$ when $p_i$ reconnects to the network at time
  $t'$. Since $p_i$ has the most recent status about $p_f$, then, due
  to Lemma~\ref{propagation_lemma_mobility}, this information about
  the mistake of $p_f$ will be propagated to all correct nodes in the
  network.  Nonetheless, as soon as $p_f$ is faulty, due to the same
  arguments of Lemma~\ref{L-strong-complet} and
  Lemma~\ref{propagation_lemma_mobility}, $p_f$ is permanently
  included in every $suspected$ set of a correct process, either
  moving or non-moving.

\end{proof}

\begin{lemma}
\label{L-mobil-weak-accuracy}
Consider an {\it f-covering network}  in which all nodes satisfy $\mathcal{MP}$ and all moving nodes satisfy  $\mathcal{M}obi\mathcal{P}$.
Let $p_i$ be a correct non-moving node which satisfies the mobility responsiveness property  $\mathcal{M}obi\mathcal{RP} (p_i)$. There is a time $u$ after which $p_i$ is not included in the $suspected_j$ set of any correct process $p_j$ (moving or non-moving).
\end{lemma}
\begin{proof}


Since  $\mathcal{M}obi\mathcal{RP} (p_i)$ is satisfied, there is a time $s$ after
which,  $\mathcal{RP} (p_i)$ holds and nodes in the neighborhood of $p_i$ do not leave $range_i$. 
Thus, due Lemma~\ref{L-weak-accuracy} (Remark 1), there is a time $s'$ after
which, no process in the network adds $p_i$ to its suspected set
(on to the execution of lines~\ref{new_susp}--\ref{end_new_susp}).

 Due to Lemma~\ref{L-weak-accuracy} (Remark 2), we can ensure that $p_i$ will
  not be included in any suspected set of {\em non-moving} correct nodes. We
  must then prove that eventually $p_i$ is not included in the
  $suspected_m$ set of any correct {\em moving} node $p_m$.
  Let us consider a correct moving node $p_m$ starting to move at time $t$ and stopping to move at time $t'$. Notice that, if $p_m$ does not suspect $p_i$ before moving at time $t$, the claim follows from Lemma~\ref{L-weak-accuracy} (Remark 2). Suppose that $p_m$ suspects $p_i$ before or at time $t$. Then, since $p_m$ keeps its state during the moving, $p_i \in suspected_m$ when $p_m$ reconnects to the network at time $t'$.
  If the suspicion over $p_i$ represents the most recent information in the network, due to
  Lemma~\ref{propagation_lemma_mobility}, it is going to be diffused to all correct nodes.
  Nonetheless, as soon as $p_i$ is correct, $p_i$ will revoke such a suspicion by the execution of lines \ref{new_mistake}-\ref{end_new_mistake}, which will generate a new mistake with a greater tag. Due to Lemma~\ref{propagation_lemma_mobility}, this mistake will be propagated to all correct processes, then $p_m$ will permanently remove $p_i$ from its $suspected_m$ set.

\end{proof}

\begin{theorem}
\label{L-strong-class} Algorithm~\ref{alg:FD_async_mob} implements a
failure detector of class $\diamondsuit S$, assuming an {\it f-covering network} of moving and non-moving nodes which satisfies the behavioral properties $\mathcal{RP}$, $\mathcal{MP}$, $\mathcal{M}obi\mathcal{P}$ and $\mathcal{M}obi\mathcal{RP}$.
\end{theorem}
\begin{proof}
The {\small\sf strong completeness} property follows directly from Lemma~\ref{L-mobil-strong-complet}.
The {\small\sf eventual weak accuracy} property follows directly from Lemma~\ref{L-mobil-weak-accuracy} and the theorem follows.
\end{proof}


\section{Performance Evaluation}
\label{perf}

In this section we study and evaluate the behavior of our
asynchronous failure detector compared to a timer-based one. To this
end, we have chosen the gossip-based heartbeat unreliable failure
detector proposed by Friedman and Tcharny  in \cite{fri05}.

Our performance experiments were conducted on top of the OMNeT++
discrete event simulator \cite{omnet:url}. We assume a
two-dimensional region $S$ of $700m$x$700m$. Transmission range $r$
is set to $100m$ in all runs. The number of nodes $N$ is fixed to
100 and  each simulation lasts 30 minutes. The one-hop network delay
$\delta$ is equal to $1ms$ in average. Since our unreliable failure
detector needs a network where the $f\_covering$ property always
holds, the $N$ nodes can not be placed randomly inside the region
$S$. The initial topology of the network is in fact gradually built
before the beginning of execution of an experiment. Thus, we start
by inserting a graph clique of $f+2$ nodes organized in a circle
whose radius is equal to $r/2$. Then, at each step, a new node of
$S$ is randomly chosen. The latter is included in the network
regardless it has $f+1$ neighbors in the current configuration. The
construction of the network stops when it reaches $N$ nodes.

In the unreliable FD proposed by Friedman and Tcharny, a node
periodically sends heartbeat messages to its neighbors. A vector is
included in every {\it heartbeat} message such that each entry in
the vector corresponds to the highest {\it heartbeat} known to be
sent from the corresponding node. Every $\Delta$ time units, each
node increments the entry of the vector corresponding to itself and
then broadcasts its {\it heartbeat} to its neighbors. Based on the
performance experiments described in the authors's article, we have
set $\Delta$ to $1s$. Upon receiving a {\it heartbeat} message, a
node updates its vector to the maximum of its local vector and the
one included in the message. A node also associates a timer to each
other node of the system. Thus, node $j$ set the timer of $i$ to
$\Theta$ whenever it receives a new information about $i$. On the
other hand, if the timeout of $i$ expires, it is considered
suspected by $j$. Note that the value of $\Theta$ should take into
count higher communication delay due to longer paths between two
nodes. We have set the value of $\Theta$ to 2s.

Concerning the implementation of our FD, it is not feasible that a
node continuously broadcasts a {\sc query}  message since the
network would be overloaded with messages. To overcome this problem,
we have included a delay of $\Delta$ units of time between lines
\ref{receive} and \ref{recfrom} of the Algorithm \ref{alg:FD_async}.
Similar to the Friedman and Tcharny's approach, we have set $\Delta$
to $1s$. However, by adding this waiting period, a processes may
receive more than $d-f$ replies. Therefore, the extra replies will
also be included in the $rec\_from$ set of this process (line
\ref{recfrom}), reducing then the number of false suspicions. It is
worth remarking that this improvement does not change the protocol
correctness.

\subsection{Failure Detection}
\label{completness}

\begin{figure}[htbp]
\begin{center}
\includegraphics[scale=0.6]{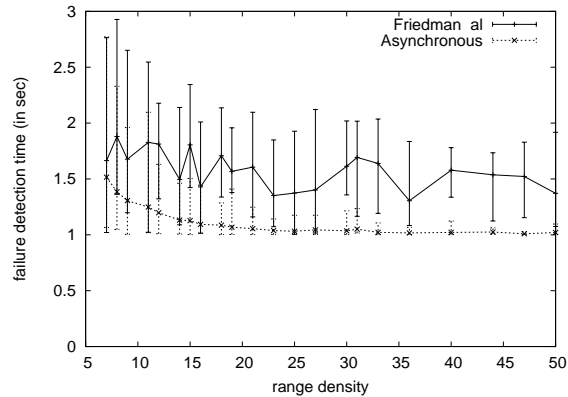}
\end{center}
\vspace{-0,3cm}
\caption{Failure detection time vs. density}
\label{fig:completude}
\end{figure}

In order to evaluate the completeness property of both failure
detectors, we have measured the impact of the range density $d$ of
the network on their respective failure detection time (Figure \ref
{fig:completude}). The number of faults is equal to $5$ and they are
uniformly inserted during an experiment. The range density $d$
varies from $7$ to $N/2$ nodes. For each density, we have measure
the average, maximum and minimum failure detection time.

We observe that for both failure detectors there is no false
suspicion. Furthermore, the propagation of failure suspicions is
quite fast because the diameter of the network is relatively small.
In the case of Friedman and Tcharny's FD, the mean failure detection
time is always between $\Theta - \Delta$ and $\Theta$ time units,
independently of $d$ since failures are detected based on heartbeat
vector values and timers. Such limit values can be explained: if
node $i$ crashes just after node $j$ has set its timer related to
$i$ to $\Theta$, $j$ will detect the crash of $i$ after $\Theta$
units of time; if $i$ crashes just before broadcasting a heartbeat,
i.e. just after $\Delta$ units of time, $j$ will detected the crash
of $i$ after $\Theta - \Delta$ units of time. On the other hand, for
our FD, the failure detection time decreases with the range density.
This happens because failure detection information is included in
{\sc query} messages which spreads faster over the network when the
density increases. We can notice that for values of $d$ greater than
22, the failure detection time is uniform and equals around $\Delta+
\delta$.

The maximum failure detection time characterizes the time for all
nodes to detect a failure (strong completeness). We can observe that
compared to Friedman and Tcharny's FD, this time is smaller and
homogeneous for our FD, which can be also explained by the above
mentioned propagation of failure information in {\sc query}
messages.



\subsection{Impact of mobility}

We have evaluated the accuracy property when a node  $m$ which has 7
neighbors and is located at one boundary of the network moves about
500m at a speed of 2m/s. It starts moving at time 100s. We consider
that while moving, node $m$ does not interact with the other nodes
as if it travels through a disturbance region where it can not send
or receive any message. Thus, $m$ stops executing while it moves.
Furthermore, all neighbors of $m$ must have $d-f+1$ neighbors. Such
restriction is necessary to guarantee that at least $d-f$ nodes will
reply to the query of these old neighbors of $m$ after it moves. The
range density $d$ of the network is equal to $7$ and there is no
fault.
\begin{figure}[htbp]
\begin{center}
\includegraphics[scale=0.6]{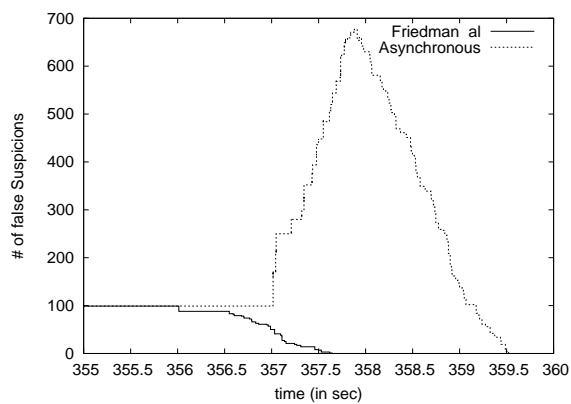}
\end{center}
\vspace{-0,3cm} \caption{Total number of false suspicions}
\label{fig:accuracy_mobility}
\end{figure}

For each experiment, the total number of false suspicions has been
measured.  Figure \ref{fig:accuracy_mobility} shows the moment just
before and after node $m$ stops moving at time 356s. We can observe
that all  $N-1$ nodes suspect $m$ before this time in both failure
detectors. After it, false suspicions about node $m$ start being
corrected by all nodes. In Friedman and Tcharny's FD, there are no
more false suspicions in around 1.5s. False suspicions about node
$m$ will also start being corrected in our FD since $m$  generates a
mistake which is propagated over the network. However, node $m$ at
the same time starts suspecting its 7 old neighbors. Thus, it
broadcasts such suspicions in its next {\sc query} message. This
information spreads over the network and nodes of the system will
start suspecting them too. This is the reason why the total number
of false suspicions starts increasing after 357s till 358s when
almost all nodes suspect the 7 old neighbors of $m$. However, at
this time such an information also reaches the latter that then
generate the corresponding mistakes and broadcast them. Such
mistakes are propagated to all nodes of the network. All false
suspicions are corrected by all nodes at 359.5s.



\section{Related Work}
\label{relwork}

As in our approach, some scalable failure detector implementations
do not require a fully connected network. Larrea {\em et al.} proposed in
\cite{larrea00} an implementation of an unreliable failure detector based on a
logical ring configuration of processes. Thus, the number of
messages is linear, but the time for propagating failure information
is quite high. In \cite{gupta01scalable}, Gupta et al. proposed a
randomized distributed failure detector algorithm which balances the
network communication load. Each process randomly chooses some
processes whose aliveness is checked. Practically, the randomization
makes the definition of timeout values difficult. In \cite{BMS03}, a
scalable hierarchical failure adapted for Grid configurations is
proposed. However, the global configuration of the network is initially known by all nodes.
It is worth remarking that none of these works tolerate mobility of nodes.

Few implementations of unreliable failure detector found in the
literature focus on MANET environments. All of them are timer-based
ones. In the Friedman and Tcharny algorithm \cite{fri05}, authors
assumes a known number of nodes and that failures include message
omissions too.
In \cite{tai04}, the authors exploit a cluster-based communication
architecture for implementing a failure detector service able to
support message losses and node failures. However, they provide
probabilistic guarantees for the accuracy and completeness
properties.

Sridhar presents in \cite{sri06} the design of a hierarchical
failure detection which consists of two independent layers: a local
one that builds a suspected list of crashed neighbors of the
corresponding node and a second one that detects mobility of nodes
across network, which  corrects possible mistakes. Contrarly to our
approach that allows the implementation of FD of class $\diamondsuit
S$, the author's failure detector is an eventually perfect {\it
local} failure detector of class $\diamondsuit P$ i.e., it provides
strong completeness and eventual strong accuracy but with regard to
a node's neighborhood.

In order to solve the problem of reaching
agreement in mobile networks where processes can crash, Cavin et al. \cite{cav05} have adapted the failure detector
definition of \cite{chandra96unreliable} to the case where the
participants are unknown.  They have introduced the concept of local participant detectors, which are
oracles that inform the subset of processes that participating in the consensus. 
The authors construct an algorithm that solves consensus with an unknown number of participants in a fail-free network. Furthermore, they extend their solution and prove that a perfect failure detector ($\mathcal{P}$) is required for solving the fault-tolerant consensus with a minimum degree of connectivity.
Greve {\em et al.}\cite{gre07} have subsequently extended this work, by providing a solution for the consensus in a fail-prone network which considers the minimal synchrony assumption (i.e., the $\diamondsuit S$), but at the expenses of requiring a higher degree of connectivity involving with the set of participants. We believe that our proposed  $\diamondsuit S$ FD will be of great interest to implement this consensus algorithm over a MANET.

\section{Conclusion}
\label{conclusion}

This paper has presented a new implementation of an unreliable
failure detector for dynamic networks such as MANETs, where the
number of nodes is not initially known and the network is not fully
connected. Our algorithm is based on a query-response mechanism
which is not timer-based. We assume that the network has the {\it
f-covering} property, where $f$ is the maximum number of failures.
This property guarantees that there is always a path between two
nodes despite of failures. Our algorithm can implement failure
detectors of class $\diamondsuit S$ when both the behavioral
responsiveness ($\mathcal{RP}$, $\mathcal{M}obi\mathcal{RP}$), membership ($\mathcal{MP}$) and mobility ($\mathcal{M}obi\mathcal{P}$) properties  are satisfied by the underlying system.
The proposed algorithm supports mobility of nodes as well.
As a future work, we plan to adapt our algorithms and properties to implement other classes of failure detectors.

\bibliographystyle{alpha}
\bibliography{biblio}

\tableofcontents

\end{document}